# Analysis of False Data Injection Impact on AI-based Solar Photovoltaic Power Generation Forecasting

Salih Sarp and Murat Kuzlu
*Old Dominion University*
*Norfolk, VA, USA*

Umit Cali
*Norwegian University of Science and Technology,*
*Trondheim, Norway*

Onur Elma
*Yildiz Technical University,*
*Istanbul, TURKEY*

Ozgur Guler
*eKare, Inc.*
*Fairfax, VA, USA*

*Abstract-* **The use of solar photovoltaics (PV) energy provides additional resources to the electric power grid. The downside of this integration is that the solar power supply is unreliable and highly dependent on the weather condition. The predictability and stability of forecasting are critical for the full utilization of solar power. This study reviews and evaluates various machine learning-based models for solar PV power generation forecasting using a public dataset. Furthermore, The root mean squared error (RMSE), mean squared error (MSE), and mean average error (MAE) metrics are used to evaluate the results. Linear Regression, Gaussian Process Regression, K-Nearest Neighbor, Decision Trees, Gradient Boosting Regression Trees, Multi-layer Perceptron, and Support Vector Regression algorithms are assessed. Their responses against false data injection attacks are also investigated. The Multi-layer Perceptron Regression method shows robust prediction on both regular and noise injected datasets over other methods.**

## I. INTRODUCTION

A reliable power system is required to maintain grid stability and security. Considering that the electric energy cannot be stored in large quantities, the real-time power supply and load balance are key factors for the power system's stable operation. Power grids are undergoing rapid change with increased penetration of Distributed Energy Resources (DER), i.e., wind and solar photovoltaics (PV). Especially solar PV has shown unprecedented growth in the U.S. during recent decades. The integration of renewable energy technologies into the existing grid requires robust performing forecasting algorithms due to their weather dependency. At present, there are more than 1 million solar PV installations across the U.S., representing 71.3 GW of operating PV capacity. The increase in residential installations helped the U.S. solar market grow 45% year-over-year [1]. In Q3 2019, solar capacity was significantly increased when compared to other fuel types, i.e., accounting for 39% of all new electric generating technologies [2]. The availability of solar energy creates an alternative source and enhances the reliability of the energy supply. However, the grid integration of weather-dependent power generation technologies, such as wind and PV, has its challenges due to fluctuations in power generation [3]. Energy forecasting using machine learning (ML) is one of the most promising techniques to increase the grid integration capabilities of wind and PV resources [4, 5]. Artificial Intelligence (AI) methods have achieved competitive prediction performance due to their success in extracting the complex underlying structure of the solar data [6]. AI methods, especially deep learning (DL) algorithms, can be implemented without feature engineering and are less sensitive to missing data [7]. ML techniques, including Linear Regression (LR), K-Nearest Neighbor (KNN), Decision Trees (DT), Gaussian Process Regression (GPR), Gradient Boosting Regression Trees (GBRT), Support Vector Regression (SVR), and Multi-layer Perceptron Regression (MLPR), are used for many forecasting tasks. Prior efforts for forecasting PV power generation include developing a power generation prediction system using wavelet transformation and AI combinations [8]. SolarNet, a deep convolutional neural network (CNN) model, is proposed for solar radiation forecasting in [9]. A CNN framework for the solar prediction model based on weather data is proposed in [10]. The study in [11] uses the long short-term memory (LSTM) model for the solar power generation prediction with the utilization of the principal component analysis (PCA) to reduce the training time and improve the generalization ability of the model. The authors in [12] review time series and Artificial Neural Networks (ANN) for short-term PV power generation forecasting across five different sites. The authors in [13] review recent AI applications on solar power generation forecasting, focusing on DL techniques. The study [14] provides a PV power forecasting model using weather classification in combination with ANN using an additional aerosol index feature. A Random Forest algorithm optimized by Differential Evolution Grey Wolf Optimizer [15] is used to forecast the PV power generation. The authors in [16] proposed the hour-ahead PV power generation forecast using SVR, Polynomial Regression, and Lasso. The study [17] summarizes the solar power generation forecasting performance of ANN, SVR, and GPR with sensitivity analysis for parameter tuning. This study attempts following contributions to the research field:

(i) The development of an AI-based PV power generation forecasting model,
(ii) Validation of the model with RMSE, MSE, and MAE metrics,
(iii) Analyses of noise impact on ML models.

## II. MACHINE LEARNING TOOLS

Several machine learning techniques are used effectively in solar power generation forecasting. The current study focuses on the algorithms, LR, GPR, KNN, DT, GBRT, and SVR algorithms with the goal to find a robust method for solar power generation forecasting. Other regression models, such as linear regression with Stochastic Gradient Descent (SGD), Bayesian Ridge, Lasso with Least Angle Regression, Automatic Relevance Determination (ARD), Passive Aggressive (PA), and Theil-Sen regressor models have also been experimented within this study.

The general overview of this study is shown in Fig. 1. In the first part of this study, the dataset without noise is used for forecasting, and results are evaluated with performance metrics. In the second part, noise is added to the testing dataset, and the performance of each method is compared using the RMSE metric.

### A. Linear Regression

LR is used to extract the relationship between input variables and an output variable. The math and statistics behind the LR are well defined, and the LR model computes a weighted sum of the input features and a bias term using techniques such as least squares, gradient descent, and regularizations [18].

$$y' = \theta_0 + \theta_1 x_1 + \theta_2 x_2 + \theta_3 x_3 + \cdots + \theta_n x_n$$

where y' is the predicted value, the number of features is n, $x_i$ is the i[th] feature variable, $\theta_j$ is the j[th] regression coefficient including the bias term.

### B. Gaussian Process Regression

GPR is a non-parametric Bayesian probabilistic regression that approximates the target value as a posterior distribution by combining a prior and a likelihood (Gaussian noise) model [19].

### C. K-Nearest Neighbor Regression

KNN is based on a feature distance/similarity measurement on data pairs to predict the numerical target [20]. This paper used the number of neighbors as two and Euclidean distance to assign the average value of the neighbors.

### D. Decision Trees

DT takes a set of features as input and carries out multiple binary classifications that provide a correlation between the input data and output response.

### E. Gradient Boosting Regression Trees

The GBRT technique corrects the mistakes of prior base models and utilizes additive training (boosting), which combines the additional trees to improve the prediction accuracy. Extreme Gradient Boosting was used for this study to achieve higher performance.

### F. Support Vector Regression

The SVR method maps the inputs into a high-dimensional feature space by constructing a decision surface. The relationship between inputs and outputs is extracted via mapping operation. The linear regression is employed on this high dimensional space for forecasting.

### G. Multi-Layer Perception Neural Network Regression

The MLPR is a feedforward ANN that consists of 100 hidden layers, ReLU activation function, 500 max number of iterations, and 0.001 learning rate with Adam optimizer.

## III. DATA COLLECTION AND PREPROCESSING

### A. Dataset Collection

An open-source benchmark dataset is used for this work, which is obtained from the Global Energy Forecasting Competition (GEFCOM) held in 2014. The dataset consists of hourly PV power generation data and corresponding numerical weather forecasts from April 2012 to July 2014, and contains the following 12 weather variables from the European Centre for Medium-Range Weather Forecasts (ECMWF):

1) Total column liquid water (*kg m\*\*-2*)
2) Total column ice water (*kg m\*\*-2*)
3) Surface pressure (*Pa*)
4) Relative humidity (*%*)
5) Total cloud cover (0-1)
6) 10-meter U wind component (*m s\*\*-1*)
7) 10-meter V wind component (*m s\*\*-1*)
8) 2-meter temperature (*K*)
9) Surface solar rad down (*J m-2*)
10) Surface thermal rad down (*J m-2*)
11) Top net solar rad (*J m-2*)
12) Total precipitation (*m*)

### B. Train/Test Dataset and Validation

The dataset is split into training and testing subsets with a ratio of 0.8 and 0.2 accordingly. The randomized split is seeded to obtain the same division for further validation of the model.

The RMSE metric is used to evaluate and compare different ML models. The RMSE metric gives the same level of error with the prediction, which makes the interpretation easier. Moreover, the RMSE metric gives smaller values, which are widely preferred for simplicity. The RMSE equation is shown in equation (1). Another metric that is used in this study is MAE, which quantifies the accuracy of our model by calculating the average absolute difference between actual and predicted values in equation (2).

$$RMSE = \sqrt{\frac{\sum(Y_t - \hat{Y}_t)^2}{n}} \quad (1)$$

$$MAE = \frac{1}{n}\sum_{t=0}^{n}|Y_t - \hat{Y}_t| \quad (2)$$

where :
$Y_t$ : The actual t[th] instance
$\hat{Y}_t$ : The forecasted t[th] instance
n: The total number of testing instance

The MSE scores are utilized for further analyses of the model using the same calculation as in Eq. (1), without a square root of the overall computation. MAE also has the same unit as the prediction.

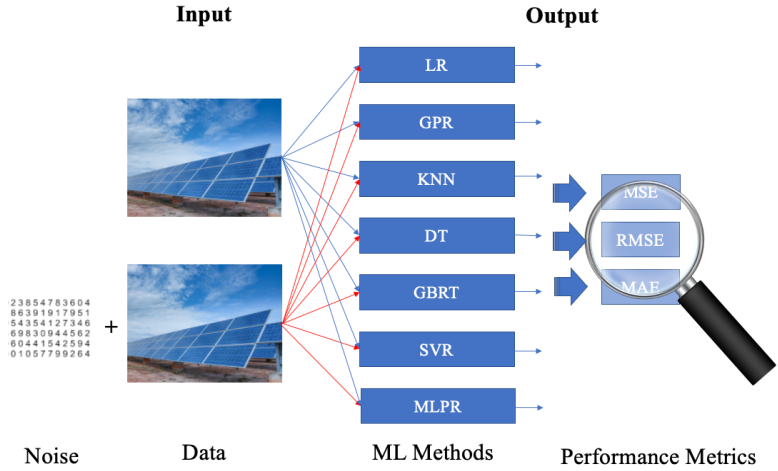

Figure 1. The framework of the solar PV power generation forecasting model

## IV. IMPLEMENTATION OF PV GENERATION FORECASTING AND VALIDATION USING DIFFERENT ML TECHNIQUES

The objective of this paper is to employ different ML techniques to evaluate and analyze the performance of each technique on a regular and noise-injected dataset so that the PV power generation will be broadly utilized in smart grid applications with higher acceptance. Each model is trained using the training set, and the performance is evaluated through the test set. The variation between predicted and actual values is compared using different metrics. In the second run, the dataset was modified with Gaussian noise, and the ML models are evaluated according to their noise sensitivity using the RMSE metric.

### A. PV Power Generation Forecasting

The PV power generation forecasting has critical importance as the integration of solar panels into the power grid depends on stable and predictable power generation. This study provides a comparison of different machine learning techniques needed for dependable power generation forecasting. The performance of each technique is calculated and shown in Table I. The lower scores are better. The graphical prediction representations of these models are depicted in Fig. 2.

The performance of linear models suffers from higher error scores, where the Lasso algorithm has the highest error rates of RMSE (0.2601), MSE (0.0676), and MAE (0.2172) scores. The modified Linear regression models, i.e., Passive Aggressive, Theil-Sen, SGD, ARD, Bayesian Ridge regression methods, have higher RMSE, MSE, and MAE scores similar to the regular LR model. Thus, these modified regression models are omitted from the rest of the study.

The performance of the DT regression model is also low because it suffers from working with continuous variables. SVR and MLPR models provide the highest performance in terms of RMSE and MSE scores. On the other hand, MAE scores of GPR and KNN are the lowest among the ML methods. RMSE and MSE scores of GPR and KNN methods are close to SVR as well.

We can conclude that these methods fit the model very closely. The performance of the GBRT method indicates that it is the third successful model for having low scores of RMSE, MSE, and MAE.

Table I.
The performance of each ML technique

| Metric Technique | RMSE | MSE | MAE |
|---|---|---|---|
| LR | 0.1261 | 0.0159 | 0.0901 |
| GPR | 0.1034 | 0.0107 | 0.0541 |
| KNN | 0.1035 | 0.0107 | 0.0509 |
| DT | 0.1351 | 0.0182 | 0.0628 |
| GBRT | 0.1044 | 0.0109 | 0.0637 |
| SVR | 0.0923 | 0.0085 | 0.0628 |
| MLPR | 0.0959 | 0.0091 | 0.0599 |
| SGD | 0.1263 | 0.0159 | 0.0910 |
| Bayesian Ridge | 0.1262 | 0.0159 | 0.0901 |
| Lasso | 0.2601 | 0.0676 | 0.2172 |
| ARD | 0.1266 | 0.0160 | 0.0901 |
| PA | 0.1603 | 0.0256 | 0.1282 |
| Theil-Sen | 0.1331 | 0.0177 | 0.0887 |

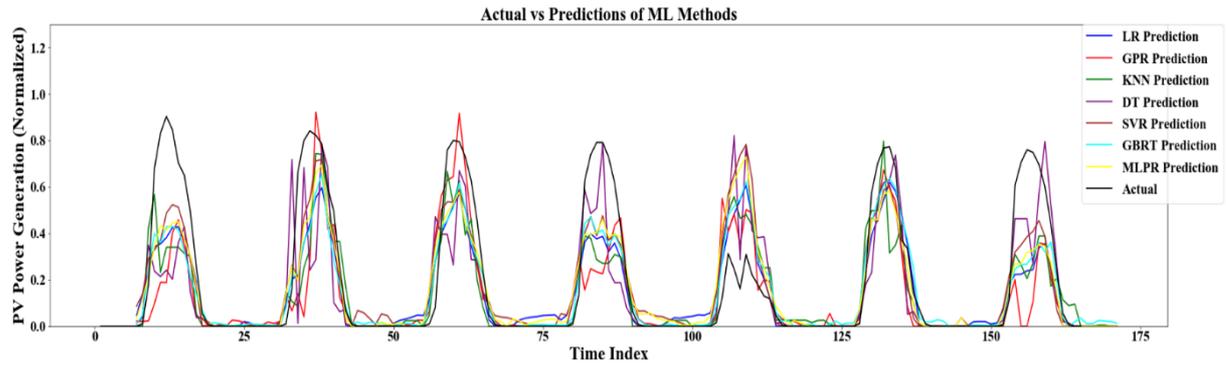

Fig. 2: Comparison of different ML techniques for forecasting.

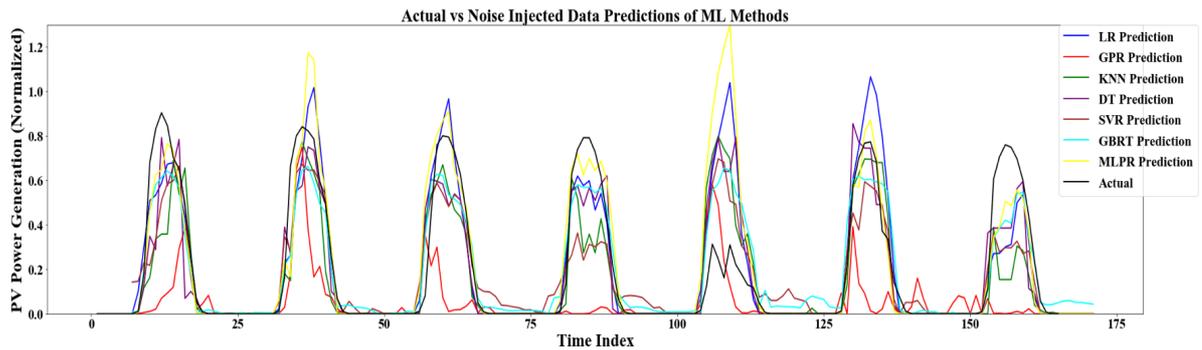

Fig. 3: Comparison of different ML techniques using noise injected dataset.

*B. The Comparison Of Machine Learning Techniques On Noise-Injected Dataset*

The impact of noise on PV power generation forecasting is important to evaluate the robustness of the forecasting models [21]. The collected data comes from various sources using many sensors, which are exposed to adversarial attacks and noise. The performance of each model is evaluated through RMSE, MAE, and MSE metrics for a noisy dataset. Only the RMSE scores of the experiment are given for simplicity.

Gaussian noise is produced using the Scikit-learn library. The injected noise has a standard deviation equal to one and a mean equal to zero. Noise is applied to a subset of the dataset, i.e., to 10%, in the first run, 50% in the second run, and 100% (all data) in the third run. The RMSE scores of the experiments are shown in Table II.

Table II.
RMSE scores of different ML techniques with noise injected dataset.

| Percentage Technique | 0% | 10% | 50% | 100% |
|---|---|---|---|---|
| LR | 0.1261 | 0.1261 | 0.1261 | 0.1261 |
| GPR | 0.1034 | 0.1035 | 0.1034 | 0.1036 |
| KNN | 0.1035 | 0.1036 | 0.1031 | 0.1027 |
| DT | 0.1351 | 0.1350 | 0.1354 | 0.1364 |
| GBRT | 0.1044 | 0.1044 | 0.1046 | 0.1047 |
| SVR | 0.0923 | 0.0922 | 0.0922 | 0.0924 |
| MLPR | 0.0959 | 0.0959 | 0.0958 | 0.0959 |

According to the results, the model performance reduces with a higher noise injection for some models. LR and MLPR show a very small change to the noise-injected dataset. The RMSE scores of SVR, GBRT, and GPR indicated that these models are sensitive to noise. On the other hand, the RMSE score of the KNN method is reduced with the increasing noise level. This will be a result of overfitting. The RMSE score of the DT is increased considerably with noise as well. The forecasting of different ML methods on the noise-injected dataset is shown in Fig. 3.

*V. RESULTS AND DISCUSSION*

The results are indicating that most of the ML techniques provide reasonable results for forecasting solar PV power generation. RMSE, MAE, and MSE metrics correlate as expected. The least successful models for the regular dataset are DT, LR, and GBRT models, which have RMSE values of 0.1351, 0.1261, and 0.1044, respectively. Forecasting with other linear models produced higher error scores on a regular dataset. SVR and MLPR methods have the lowest RMSE scores, i.e., 0.0923 and 0.0959. GPR and KNN have similar RMSE scores, i.e., 0.1034 and 0.1035, but the MAE score of KNN is the lowest among ML techniques, i.e., 0.0509. GPR has a lower MAE score, i.e., 0.0107, than KNN. GBRT has slightly higher RMSE, MSE, and MAE scores., i.e., 0.1034, 0.0107, and 0.0541, respectively.

The RMSE score changes of ML methods with noise injection (see Table III) indicate that MLPR and LR are not sensitive to noise injection, 0% change in terms of RMSE. A decrease in the RMSE score for KNN shows potential overfitting. DT and GBRT indicate a high noise sensitivity, i.e., 0.21% for DT and 0.29% for GBRT, whereas GPR and MLPR show a slightly lower sensitivity, i.e., 0.01% for GPR and -0.02% for MLPR.

Table III.
RMSE score changes of ML methods with noise injection change.

| Diff.% Technique | 0% vs.10% | 0% vs. 50% | 0% vs. 100% |
|---|---|---|---|
| LR | 0.00% | 0.00% | 0.00% |
| GPR | 0.10% | 0.01% | 0.02% |
| KNN | 0.10% | -0.35% | -0.77% |
| DT | -0.07% | 0.21% | 0.9% |
| GBRT | 0.00% | 0.18% | 0.29% |
| SVR | -0.11% | -0.03% | 0.12% |
| MLPR | 0.00% | -0.02% | 0.00% |

RMSE, MSE, and MAE values are small due to the normalized data for both regular and noise injected datasets. Observations derived from the results of solar PV power generation forecasting and the noise impact are outlined below:

*Observation 1*: ML models have great potential to forecast solar PV power generation.
*Observation 2:* Noise injection affects the performance of the forecasting models.
*Observation 3:* SVR and MLPR models achieve the best performance for both regular and noise injected datasets.
*Observation 4:* LR, GPR, and MLPR models are less sensitive to noise.
*Observation 5*: DT and GBRT models have a higher sensitivity to noise than other ML methods.

## VI. CONCLUSION

This paper presents solar PV power generation prediction models by utilizing seven ML techniques and the impact of noise injection on solar power forecasting in terms of performance metrics. In the first part of the study, the proposed model gets the PV power generation and weather data to forecast the generated solar power. Results of each ML technique are evaluated through RMSE, MSE, and MAE performance metrics. The SVR and MLPR models achieve better PV power generation forecasting accuracy. In the second part of the study, the impact of noise injection on the model performance is studied. The noise is injected into the testing data subset with different percentages, i.e., 10%, 50%, and 100%, and the performance differences of seven ML methods are examined in detail. LR, GPR, and MLPR are less sensitive to noise. Furthermore, the SVR model provides the highest performance on the regular dataset, while the MLPR model performs best on the noise-injected dataset. Furthermore, this study demonstrates the sensitivity analysis based on different ML algorithms, which are used for solar forecasting and their impacts of adversarial cyber-attacks like data poisoning or noise injection on the accuracy of the prediction outputs.

This study demonstrated that ML techniques could be applied successfully to regular and noisy PV power generation forecast models. It is expected that this study is useful for developers and researchers working on PV power generation forecasting.

REFERENCES
[1] U.S. Solar Market and 15 States See Best Quarter Ever for Residential Solar, https://www.seia.org/news/us-solar-market-and-15-states-see-best-quarter-ever-residential-solar
[2] Solar Market Insight Report 2019 Q4, http://www.seia.org/research-resources/solar-market-insight-report-2019-q4
[3] Ahmed, Razin, et al. "A review and evaluation of the state-of-the-art in PV solar power forecasting: Techniques and optimization." Renewable and Sustainable Energy Reviews124 (2020): 109792.
[4] Cali, Umit, and Claudio Lima. "Energy Informatics Using the Distributed Ledger Technology and Advanced Data Analytics." Cases on Green Energy and Sustainable Development. IGI Global, 2020. 438-481.
[5] Ahmad, Tanveer, and Huanxin Chen. "Utility companies strategy for short-term energy demand forecasting using machine learning based models." Sustainable cities and society 39 (2018)
[6] Wang, Huaizhi, et al. "Taxonomy research of artificial intelligence for deterministic solar power forecasting." Energy Conversion and Management 214 (2020): 112909.
[7] Belu, Radian. "Artificial intelligence techniques for solar energy and photovoltaic applications." Handbook of Research on Solar Energy Systems and Technologies. IGI Global, 2013. 376-436.
[8] Mandal, Paras, et al. "Forecasting power output of solar photovoltaic system using wavelet transform and artificial intelligence techniques." Procedia Computer Science 12 (2012): 332-337.
[9] Kuo, Ping-Huan, and Chiou-Jye Huang. "A green energy application in energy management systems by an artificial intelligence-based solar radiation forecasting model." Energies 11.4 (2018): 819.
[10] Dong, N., Chang, J.F., Wu, A.G. and Gao, Z.K., 2020. A novel convolutional neural network framework based solar irradiance prediction method. International Journal of Electrical Power & Energy Systems, 114, p.105411.
[11] J. Zhang, Y. Chi and L. Xiao, "Solar Power Generation Forecast Based on LSTM," 2018 IEEE 9th International Conference on Software Engineering and Service Science (ICSESS), Beijing, China, 2018, pp. 869-872.
[12] Isaksson, E. and Karpe Conde, M., 2018. Solar Power Forecasting with Machine Learning Techniques.

[13] Mellit, Adel, et al. "Advanced Methods for Photovoltaic Output Power Forecasting: A Review." Applied Sciences 10.2 (2020): 487.

[14] J. Kou et al., "Photovoltaic power forecasting based on artificial neural network and meteorological data," 2013 IEEE International Conference of IEEE Region 10 (TENCON 2013), Xi'an, 2013, pp. 1-4,

[15] Liu, D. and Sun, K., 2019. Random forest solar power forecast based on classification optimization. Energy, 187, p.115940.

[16] A. Alfadda, R. Adhikari, M. Kuzlu and S. Rahman, "Hour-ahead solar PV power forecasting using SVR based approach," 2017 IEEE Power & Energy Society Innovative Smart Grid Technologies Conference (ISGT), Washington, DC, 2017, pp. 1-5.

[17] Sharifzadeh, Mahdi, Alexandra Sikinioti-Lock, and Nilay Shah. "Machine-learning methods for integrated renewable power generation: A comparative study of artificial neural networks, support vector regression, and Gaussian Process Regression." Renewable and Sustainable Energy Reviews 108 (2019): 513-538.

[18] Géron, Aurélien. Hands-on machine learning with Scikit-Learn, Keras, and TensorFlow: Concepts, tools, and techniques to build intelligent systems. O'Reilly Media, 2019.

[19] Park, Jinkyoo, et al. "Gaussian Process Regression (GPR) Representation in Predictive Model Markup Language (PMML)." Smart and sustainable manufacturing systems 1.1 (2017): 121.

[20] Khamar, Khushbu. "Short text classification using kNN based on distance function." International Journal of advanced research in computer and communication engineering 2.4 (2013): 1916-1919.

[21] Singh, Sameer. "Noise impact on time-series forecasting using an intelligent pattern matching technique." Pattern Recognition 32.8 (1999): 1389-1398.